\documentclass[preprint, showpacs, aps, draft]{revtex4}
\usepackage{bm}
\usepackage{amsthm}

\begin{document}
\title{Maximum observable correlation for a bipartite quantum system}
\author{Michael J. W. Hall}
\affiliation{Theoretical Physics, IAS,  Australian National
University,
Canberra ACT 0200, Australia}

\author{Erika Andersson}
\author{Thomas Brougham}
\affiliation{Department of Physics, University of Strathclyde, Glasgow G4 0NG, United Kingdom}

%\date{\today}

% USE FOR BOTH
\begin{abstract}
The maximum observable correlation between the two components of a bipartite quantum system is a property of the joint density operator, and is achieved by making particular measurements on the respective components.  For pure states it corresponds to making measurements diagonal in a corresponding Schmidt basis.  More generally, it is shown that the maximum correlation may be characterised in terms of a `correlation basis' for the joint density operator, which defines the corresponding (nondegenerate) optimal measurements.  The maximum coincidence rate for spin measurements on two-qubit systems is determined to be $(1+s)/2$, where $s$ is the spectral norm of the spin correlation matrix, and upper bounds are obtained for $n$-valued measurements on general bipartite systems.  It is shown that the maximum coincidence rate is never greater than the computable cross norm measure of entanglement, and a much tighter upper bound is conjectured. Connections with optimal state discrimination and entanglement bounds are briefly discussed.
\end{abstract}

%USE FOR REVTEX
\maketitle

\section{Introduction}

Suppose that two observers, Alice and Bob, have access to the respective components of a bipartite quantum system.  If the observers make measurements of observables $A$ and $B$ respectively, the correlation between the measurement outcomes will clearly depend on $A$ and $B$.  It is therefore of interest to ask {\it what choice of $A$ and $B$ will give the maximum possible correlation}.  The answer would allow bipartite states to be ranked in terms of their joint-correlation properties.  It is also relevant to the efficient generation of secure keys in quantum cryptography where, all other things being equal, Alice and Bob should aim to compare measurement outcomes which are maximally correlated for a given shared state \cite{crypt, schumacher}.

It is important to make a distinction here between trivial and non-trivial correlations. For example, if Alice and Bob each simply measure the unit operator, their results will of course be perfectly (but trivially) correlated.  Hence the answer to the above question is only of interest if it can be ensured that the measurement outcomes for each component have some useful degree of randomness.  This is critical, for example, if Alice and Bob wish to generate a secure cryptographic key \cite{schumacher}.  As will be shown, a natural approach is to require that the measured observables are `maximally informative' or `nondegenerate'.  This is equivalent to requiring the observables to be described by {\it maximal} probability operator measures (POMs), i.e., $A\equiv\{ |a_j\rangle\langle a_j|\}$, $B\equiv\{|b_k\rangle\langle b_k|\}$.  It turns out that this requirement is in fact naturally built into some measures of correlation (eg, the mutual information), while it must be imposed explicitly for others (eg, the coincidence rate).

For the case of a  pure bipartite state, $|\psi\rangle\langle\psi|$, there is an intuitively obvious answer to the above question:  Alice and Bob should choose $A$ and $B$ such that the kets $\{|a_j\rangle\}$ and $\{|b_j\rangle\}$ correspond to a Schmidt decomposition of $|\psi\rangle$, i.e., such that
\begin{equation} \label{schmidt}
 |\psi\rangle = \sum_j \sqrt{p_j}\, |a_j\rangle\otimes |b_j\rangle .
\end{equation}
Thus, each possible measurement outcome $A=a_j$ will be perfectly correlated with the corresponding measurement outcome $B=b_j$.  Note for this case that $\langle a_j|a_k\rangle = \delta_{jk} = \langle b_j|b_k\rangle$. Hence, the optimal observables are described by {\it orthogonal} POMs, and can be equivalently represented by the Hermitian operators $\hat{A}=\sum_j a_j |a_j\rangle\langle a_j|$ and $\hat{B}=\sum_j b_j|b_j\rangle\langle b_j|$ acting on the respective Hilbert space components \cite{helstrom,holevo}.

More generally, when the bipartite state is described by some density operator $\rho$, finding the maximal POMs $A\equiv\{ |a_j\rangle\langle a_j|\}$, $B\equiv\{|b_k\rangle\langle b_k|\}$ that maximise a given measure of correlation is quite difficult.  Such a pair of maximally-correlated observables will determine a corresponding basis set, $\{|a_j\rangle\otimes |b_k\rangle\}$, for the bipartite system.  This basis set generalises the notion of the Schmidt basis for pure states, and may be called a {\it correlation basis} for $\rho$.  Unlike the Schmidt basis, the correlation basis need not always be orthonormal.

Mutual information and coincidence rate, as measures of correlation, are briefly discussed in Sec.~II.  Formal equations for the correlation basis are given in Sec.~III, for the case of coincidence rate, and illustrated with examples in Sec.~III~C, including connections with the problem of optimal state discrimination.  It is conjectured that at least one of the optimal observables $A$ and $B$ can always be chosen to correspond to an orthogonal POM.  In Sec.~IV, the maximum coincidence rate for two-valued measurements on pairs of qubits is explicitly determined as a simple function of the spectral norm of the $3\times3$ spin correlation matrix.  This result is generalised in Sec.~V, where general upper bounds for coincidence rate are obtained for $n$-valued measurements, based on a singular value decomposition of the Fano form of the density matrix \cite{fano,fanoz}. These bounds are related to the computable cross norm \cite{ccn}, and are generalised in Sec.~VI to connect other linear correlation bounds (such as spin covariance) with entanglement properties.

\section{Mutual information vs coincidence rate}

To find the optimally correlated observables for a given bipartite system, it is necessary to first quantify joint correlation in some manner.  Now, the statistics of any two observables $A$ and $B$, measured on the respective components of the system, can always be described by corresponding probability operator measures (POMs) $\{ A_j\}$ and $\{ B_k\}$ (i.e., sets of positive operators which sum to the unit operator \cite{helstrom, holevo}), with the joint probability of measurement outcomes $A=a_j$ and $B=b_k$ for a bipartite density operator $\rho$ being given by
\[	p_{jk} = {\rm tr}[\rho A_j\otimes B_k] . \]
Any measure of correlation will be some function of the probability distribution $p_{jk}$, and two well known examples are discussed in the following.

First, the {\it mutual information} is defined by \cite{shannon}
\begin{equation} \label{mutinf}
I(A,B|\rho) := \sum_{jk} p_{jk} \log_2 \frac{p_jq_k}{p_{jk}} ,
\end{equation}
where $p_j$ and $q_k$ denote the marginal distributions for $A$ and $B$ respectively. This quantity vanishes for uncorrelated observables; is invariant under relabellings of measurement outcomes; and has a simple physical interpretation: if $A$ and $B$ are each measured for a large number of copies of $\rho$, then $I(A,B|\rho)$ is the average amount of data gained per measurement outcome of $A$, about the corresponding sequence of measurement outcomes of $B$ (as quantified by the number of bits required to represent the data), and vice versa \cite{shannon}.   

The convexity of mutual information implies that the {\it maximum} mutual information for a given state $\rho$ (also called the accessible information), can always be achieved via observables described by {\it maximal} POMs \cite{davies}, i.e., with $A_j \equiv |a_j\rangle\langle a_j|$, $B_k \equiv |b_k\rangle\langle b_k|$.  Thus,
\begin{equation} \label{imax}
I_{max}(\rho):=\max_{A,B} I(A,B|\rho) = \max_{A,B{\rm ~maximal}} I(A,B|\rho) . 
\end{equation}
While it is very difficult to determine the optimal observables $A$ and $B$ in Eq.~(\ref{imax}), a useful upper bound follows from application of the Holevo bound to the ensemble of states induced on one component of the bipartite system by a measurement on the other component (see Eqs.~(12) of Ref.~\cite{hallpra}):
\begin{equation} \label{hbound}
I_{max}(\rho) \leq \min \{ S(\rho_1), S(\rho_2) \} . 
\end{equation}
Here $S(\cdot)$ denotes the von Neumann entropy, and $\rho_1$ and $\rho_2$ are the reduced operators for the first and second components of the bipartite system.  This bound is sufficiently strong to obtain the maximum mutual information for any mixture $\rho=\sum_\alpha \lambda_\alpha |\psi_\alpha\rangle \langle \psi_\alpha|$ of pure states sharing a common Schmidt basis up to trivial phase factors, i.e., with
\[ |\psi_\alpha\rangle = \sum_j \sqrt{p^{^{(\alpha)}}_j} \exp [i\phi^{^{(\alpha)}}_j]\, |a_j\rangle \otimes |b_j\rangle  .\]
In particular, Eq.~(\ref{hbound}) is saturated by choosing $A$ and $B$ to be the maximal POMs generated by this Schmidt basis, yielding
\begin{equation} \label{imaxs}
I_{max}(\sum_\alpha \lambda_\alpha |\psi_\alpha\rangle\langle\psi_\alpha|) = -\sum_j P_j \log_2 P_j , 
\end{equation}
where $P_j:=\sum_\alpha \lambda_\alpha p^{^{(\alpha)}}_j$.  Note that for a {\it pure} state, $|\psi\rangle\langle\psi|$, the maximally correlated observables are therefore those corresponding to a Schmidt basis for $|\psi\rangle$, justifying the intuitive answer given in the Introduction.

Second, the {\it coincidence rate} measure of correlation is defined by
\begin{equation} \label{coinc}
C(A,B|\rho) := \sum_j p_{jj} .
\end{equation} 
This quantity is simply the probability of the observers obtaining matched outcomes, and reaches a maximum of unity only when the outcomes are perfectly correlated (i.e., $p_{jk}=p_j\delta_{jk}$).  It is also a little more tractable than mutual information, and will therefore be the focus of this paper. Note that coincidence rate (unlike mutual information) has no clear meaning for continuously-valued outcomes: the quantity $C=\int dx\,p_{xx}$ is not invariant under relabellings of the outcomes (eg, for $x\rightarrow\lambda x$ one has $C\rightarrow C/\lambda$ ).  Hence, only discretely-valued POMs will be considered in what follows.

Unlike mutual information, the coincidence rate does {\it not} intrinsically distinguish between trivial and non-trivial correlations.  For example, if Alice and Bob each merely measure the unit operator, they will obtain the maximum possible value of coincidence rate (unity), but the minimum possible value of mutual information (zero).  Hence, as discussed in the Introduction, it is only of interest to maximise coincidence rate subject to some constraint that ensures a useful degree of randomness for the individual measurement outcomes.  
One reasonable constraint is the requirement that {\it the measured observables are maximal POMs}.  This constraint is consistent with Eq.~(\ref{imax}) for mutual information; does not allow the observers to remove potential information about correlations by merging measurement outcomes; and automatically rules out trivial correlations.  The relevant problem of interest is then the determination of observables $A$ and $B$ which achieve the maximum value
\begin{equation} \label{cmax}
C_{max}(\rho) := \max_{A,B{\rm ~maximal}} C(A,B|\rho)= \max_{A,B{\rm ~maximal}} \sum_j \langle a_j,b_j|\rho|a_j,b_j\rangle.
\end{equation}
In analogy to Eq.~(\ref{imaxs}) for mutual information, one finds
\[ C_{max}(\sum_\alpha \lambda_\alpha |\psi_\alpha\rangle\langle\psi_\alpha|) = \sum_j P_j = 1 ,\]
for mixtures of states sharing a common Schmidt basis, including all pure states.   From Eq.~(\ref{cmax}) one also obtains the general convexity property
\[
C_{max}\left( \lambda\rho +(1-\lambda) \sigma\right) \leq \lambda C_{max}(\rho) + (1-\lambda)C_{max}(\sigma) ,
\]
Hence, if two given observables $A$ and $B$ maximise the coincidence rate for some set of states, $S_{AB}$, then this set is convex.  Eq.~(\ref{cmax}) further implies that the maximum coincidence rate for any member of $S_{AB}$ is bounded above by the largest eigenvalue of the `coincidence operator' $K_{AB}:=\sum_j |a_j\rangle\langle a_j| \otimes |b_j\rangle\langle b_j|$, and hence that $S_{AB}$ contains a pure state if and only if this largest eigenvalue is unity. 

Finally, it may be recalled that mutual information and coincidence rate are both not only useful measures of correlation per se, but may also be used to differentiate `classical' from `quantum' correlations, via corresponding Bell inequalities.  For example, if Alice can measure either of $A$ and $\overline{A}$, and Bob can measure either of $B$ and $\overline{B}$, and it is assumed that the statistics of these four observables can be generated by some classical joint probability distribution, then from Eq.~(6.5) of Ref.~\cite{caves} one has
\[ I(A,B|\rho) + I(A,\overline{B}|\rho) + I(\overline{A},B|\rho) - I(\overline{A},\overline{B}|\rho) \leq H(A) + H(B), \]
where $H(\cdot)$ denotes the Shannon entropy, while from Eq.~(8) of Ref.~\cite{qcm} one has
\[ C(A,B|\rho) + C(A,\overline{B}|\rho) + C(\overline{A},B|\rho) - C(\overline{A},\overline{B}|\rho) \leq 2 . \]
Each of these inequalities is violated, for example, by suitable spin measurements on a singlet state. The use of correlation measures to characterise the minimum degree of entanglement present has been recently discussed in Ref.~\cite{plenio}.  Connections between correlation and entanglement bounds are obtained in Secs.~V and VI below.

\section{Maximising coincidence rate}

\subsection{Conditions for extrema}

The linearity of coincidence rate with respect to $A$ and $B$ makes it straightforward to characterise the extremal observables, as per the following proposition.  The conditions for such observables to {\it maximise} coincidence rate are less straightforward, however, and are left to the next subsection. 

{\bf Proposition 1}: {\it Necessary and sufficient conditions for maximal POMs $A\equiv\{|a_j\rangle\langle a_j|\}$ and $B\equiv\{|b_j\rangle\langle b_j|\}$ to attain an {\it extremal value} of coincidence rate, for bipartite state $\rho$, are
\begin{equation} \label{prop1a} \langle a_k,b_l|\rho|a_l,b_l\rangle = \langle a_k,b_k|\rho|a_l,b_k\rangle ,~~~~~ \langle a_k,b_l|\rho|a_k,b_k\rangle = \langle a_l,b_l|\rho|a_l,b_k\rangle 
\end{equation}
for all $k\neq l$.  Moreover, these conditions are equivalent to the existence of Hermitian operators $V$ and $W$, acting on the first and second components respectively, satisfying
\begin{equation} \label{prop1b}
\left(V-\langle b_j|\rho|b_j\rangle\right)\,|a_j\rangle = 0,~~~~~\left(W-\langle a_j|\rho|a_j\rangle \right)|b_j\rangle =0 
\end{equation}
for all $j$.  The corresponding extremal value of coincidence rate is given by }
\begin{equation}  \label{prop1c}
C(A,B|\rho) = {\rm tr}_1[V] = {\rm tr}_2[W] .
\end{equation}

\begin{proof} Consider the variational quantity
\[
J := \sum_j \langle a_j,b_j|\rho|a_j,b_j\rangle - {\rm tr}_1[V\,(\sum_j |a_j\rangle\langle a_j|-\hat{1}_1\,)\,] - {\rm tr}_2[W\,(\sum_j |b_j\rangle\langle b_j|-\hat{1}_2 \,) \,],
\]
defined for arbitrary sets of kets $\{|a_j\rangle\}$ and $\{|b_j\rangle\}$ of the same cardinality, where $V$ and $W$ are Hermitian operators that act as Lagrange multipliers for enforcing the completeness constraints 
\begin{equation} \label{complete}
\sum_j|a_j\rangle\langle a_j|=\hat{1}_1,~~~~~\sum_j|b_j\rangle\langle b_j|=\hat{1}_2.
\end{equation}
Clearly, $C_{max}(\rho)$ in Eq.~(\ref{cmax}) corresponds to the global maximum of $J$ under these constraints. Letting $J(\epsilon)$ denote $J$ evaluated under the variations $|a_j\rangle\rightarrow |a_j\rangle+\epsilon |m_j\rangle$, $|b_j\rangle\rightarrow |b_j\rangle+\epsilon |n_j\rangle$, the extremal points of $J$ correspond to the solutions of $J'(0) = 0$, i.e.,
\[ \sum_j {\rm tr}_1 \left[ (|m_j\rangle\langle a_j|+  h.c.)
(\langle b_j|\rho|b_j\rangle  -V)\right] + \sum_j {\rm tr}_2 \left[ (|n_j\rangle\langle b_j|+  h.c.)
(\langle a_j|\rho|a_j\rangle  -W)\right] =0 .\]
Choosing at most one element of the $\{|m_j\rangle, |n_j\rangle\}$ to be non-vanishing (and arbitrary) then yields Eq.~(\ref{prop1b}). Multiplying the latter on the left by $\langle a_k|$ and $\langle b_k|$ further yields
\[
\langle a_k,b_j|\rho|a_j,b_j\rangle = \langle a_k|V|a_j\rangle ,~~~~~ \langle a_j,b_k|\rho|a_j,b_j\rangle = \langle b_k|W|b_j\rangle ,
\]
and Eq.~(\ref{prop1a}) immediately follows from the requirement that $V$ and $W$ are Hermitian.  Multiplying on the right of  Eq.~(\ref{prop1b}) by $\langle a_j|$ and $\langle b_j|$, and summing over $j$, yields 
\begin{equation} \label{vw}
V = \sum_j \langle b_j|\rho|b_j\rangle\,|a_j\rangle\langle a_j|,~~~~~W= \sum_j \langle a_j|\rho|a_j\rangle\,|b_j\rangle\langle b_j| . 
\end{equation}
Taking these as {\it defining} relations conversely yields Eq.~(\ref{prop1b}) from Eq.~(\ref{prop1a}).  The trace of Eq.~(\ref{vw}) yields Eq.~(\ref{prop1c}).  
\end{proof}

Proposition 1 has a formal connection to the well known problem of distinguishing between members of a given statistical ensemble.  In particular, let $\{\rho_j;\lambda_j\}$ denote the ensemble containing state $\rho_j$ with probability $\lambda_j$.  It is known that necessary and sufficient conditions for a POM $\{ \Pi_j\}$ to optimally discriminate between members of this ensemble are \cite{helstrom}
\begin{equation} \label{helstrom} 
(\Upsilon -\lambda_j\rho_j)\Pi_j = 0,~~~~~\Upsilon\geq \lambda_j\rho_j 
\end{equation}
for all $j$, for some Hermitian operator $\Upsilon$.  The first of these conditions is equivalent to Eq.~(\ref{prop1b}) of Proposition 1, for the ensembles $\{\sigma_j;p_j\}$ and $\{\tau_j;q_j\}$ defined by
$p_j\sigma_j := \langle b_j|\rho|b_j\rangle$ and $q_j\tau_j := \langle b_j|\rho|b_j\rangle$. Further, summing this first condition over $j$ yields $\Upsilon = \sum_jp_j\sigma_j\Pi_j$, corresponding to Eq.~(\ref{vw}).  

However, there is no simple analogue of the {\it second} condition in Eq.~(\ref{helstrom}) - in particular, while the conditions 
\begin{equation} \label{vwcon}
V \geq \langle b_j|\rho|b_j\rangle,~~~~~~W \geq \langle a_j|\rho|a_j\rangle
\end{equation} 
would immediately imply that $A$ optimally discriminates between members of the ensemble $\{\sigma_j;p_j\}$, and that $B$ optimally discriminates between members of the ensemble $\{\tau_j;q_j\}$, these conditions are
{\it not} sufficient to ensure a maximum for the coincidence rate, as will be shown by explicit example in Sec.~III C.  Indeed, it is not clear that these conditions are even {\it necessary}.  

Finally, some general properties of extremal observables are worth nothing.  First, for {\it pure} states, the matrix coefficients in Eq.~(\ref{prop1a}) vanish identically for $k\neq l$, for the case where observables $A$ and $B$ correspond to the Schmidt basis decomposition in Eq.~(\ref{schmidt}), and hence these observables are extremal as expected.  Second, Eq.~(\ref{prop1a}) implies that if $A$ and $B$ are extremal for two density operators $\rho$ and $\rho'$, then they are extremal for any mixture of $\rho$ and $\rho'$.  Third, if $\rho$ is invariant under some local unitary transformation, i.e., $\rho=U_1\otimes U_2\rho U_1^\dagger\otimes U_2^\dagger$, then, for a given solution $A$ and $B$ of Eq.~(\ref{prop1a}), there will be a second solution $\overline{A}$ and $\overline{B}$, with $|\overline{a}_j\rangle=U_1^\dagger |a_j\rangle$ and $|\overline{b}_j\rangle=U_2^\dagger |b_j\rangle$. A similar symmetry holds when $\rho$ is invariant under the interchange of the two component systems.  

\subsection{Maxima and $n$-valued measurements}

The second-order variation of the quantity $J(\epsilon)$ appearing in the proof of Proposition 1 immediately yields the condition $J''(0)\leq 0$ for two extremal observables $A$ and $B$ to correspond to a local maximum of coincidence rate.  This condition is required to hold only for all kets $|m_j\rangle$ and $|n_j\rangle$ satisfying
\begin{equation} \label{first} 
\sum_j \left(|m_j\rangle\langle a_j| + |a_j\rangle\langle m_j|\right) = 0 = \sum_j \left(|n_j\rangle\langle b_j| + |b_j\rangle\langle n_j|\right) 
\end{equation}
(corresponding to the completeness constraints in Eq.~(\ref{complete}), to first order in $\epsilon$).  However, the set of such kets is not straightforward to characterise explicitly, and is dependent on the particular POMs $A$ and $B$ in question, making the condition difficult to verify in practice.  In contrast, an explicit and generic condition for $C(A,B|\rho)$ to be a local maximum is obtained in Proposition 2 below, based on the Naimark extension theorem. The restricted problem of maximising coincidence rate over $n$-valued measurements is also discussed.  

Attention will be limited to the case where $\rho$ has finite support.  In particular, if $H_1$ and $H_2$ are defined to be the Hilbert spaces spanned by the eigenstates of the reduced density operators $\rho_1:={\rm tr}_2[\rho]$, $\rho_2:={\rm tr}_1[\rho]$, then it is assumed that these Hilbert spaces are finite-dimensional, i.e., 
\begin{equation} \label{dim} 
d_1:={\rm dim}(H_1) <\infty,~~~~~~d_2 := {\rm dim}(H_2) <\infty. 
\end{equation}
Now, consider a maximal POM $A\equiv\{|a_j\rangle\langle a_j|\}$ on a $d$-dimensional Hilbert space $H$, having less than or equal to $n$ non-zero elements (hence, from Eq.~(\ref{complete}), $n\geq d$).  The Naimark extension theorem then implies there is an $n$-dimensional Hilbert space $H_n$ containing $H$ as a subspace, and a maximal {\it orthogonal} POM $X\equiv\{|x_j\rangle\langle x_j|\}$ on $H_n$ (i.e., with $\langle x_j|x_k\rangle=\delta_{jk}$), such that $|a_j\rangle = E|x_j\rangle$, where $E$ denotes the $d$-dimensional projection operator from $H_n$ to $H$ \cite{holevo, partha}.  The converse result trivially holds: any maximal orthogonal POM on $H_n$, with `eigenstates' $\{|x_j\rangle\}$, generates a maximal POM $A$ on $H$ with at most $n$ non-zero elements, defined via $|a_j\rangle:=E|x_j\rangle$.  Since all $d$-dimensional subspaces of $H_n$ are unitarily equivalent, this establishes the following Lemma:

{\bf Lemma} (Naimark extension theorem for maximal POMs):  {\it For a $d$-dimensional Hilbert space, $H$, the set of maximal POMs on $H$ having at most $n\geq d$ non-zero elements is characterised by the set of maximal orthogonal POMs on any $n$-dimensional Hilbert space $H_n$ that contains $H$ as a subspace.}  

It follows immediately, taking the limit $n\rightarrow\infty$, that the class of {\it all} maximal POMs on $H$ can be represented by the class of maximal orthogonal POMs on $H_\infty$.  Thus, the joint measurement of any two maximal POMs $A$ and $B$, on the respective components of the tensor product $H_1\otimes H_2$ spanned by $\rho$, can be represented by the measurement of two maximal orthogonal POMs $X$ and $Y$ on the respective components of the tensor product $H_\infty\otimes H_\infty$, with
\begin{equation} \label{aex}
|a_j\rangle = E|x_j\rangle, ~~~~~|b_k\rangle = F|y_k\rangle, ~~~~~(E\otimes F) \rho =\rho = \rho(E\otimes F)  , 
\end{equation}
where $E$ and $F$ denote the $d_1$ and $d_2$-dimensional projections onto $H_1$ and $H_2$ respectively.  In particular, one has 
\begin{equation} \label{cxy}
C(A,B|\rho) \equiv C(X,Y|\rho) := \sum_{j=1}^\infty \langle x_j,y_j|\rho|x_j,y_j\rangle .
\end{equation}
The advantage of this representation is that maximal orthogonal POMs on $H_\infty$ are connected by unitary transformations.  This allows one to explicitly write down the necessary and sufficient conditions for an extremal value of coincidence rate to be a local maximum, as per the following proposition.  

{\bf Proposition 2}: 
{\it Two maximal orthogonal POMs $X$ and $Y$ on $H_\infty$, and hence the corresponding maximal POMs $A$ and $B$ defined via Eq.~(\ref{aex}), generate a local maximum of coincidence rate if and only if
\begin{equation} \label{prop2a}
\langle x_k,y_l|\rho|x_l,y_l\rangle = \langle x_k,y_k|\rho|x_l,y_k\rangle ,~~~~~ \langle x_k,y_l|\rho|x_k,y_k\rangle = \langle x_l,y_l|\rho|x_l,y_k\rangle 
\end{equation}
for all $k\neq l$, and 
\begin{equation} \label{prop2b}
\sum_j \left\{ \langle x_j|M(V-\langle b_j|\rho|b_j\rangle)M|x_j\rangle + \langle y_j|N(W-\langle a_j|\rho|a_j\rangle)N|y_j\rangle + {\rm tr}\left( \rho\,[M,|x_j\rangle\langle x_j|]\otimes [N,|y_j\rangle\langle y_j|]\right) \right\} \geq 0 ,
\end{equation}
for all Hermitian operators $M$ and $N$ on $H_\infty$, where $V$ and $W$ are defined as per Eq.~(\ref{vw}).}

The proof is given in the Appendix.  Note that the first condition is equivalent to Eq.~(\ref{prop1a}) of Proposition 1 (and hence to Eq.~(\ref{prop1b}) also), as an immediate consequence of Eq.~(\ref{aex}).  Further, the second condition is equivalent to the condition $J''(0)\leq 0$ discussed above, if one defines $|m_j\rangle:=iEM|x_j\rangle$ and $|n_j\rangle :=iFN|y_j\rangle$ (the constraints in Eq.~(\ref{first}) follow from the anti-Hermiticity of the operators $iEME$ and $iFNF$).  Note that the presence of the last term in Eq.~(\ref{prop2b}) implies that the conditions in Eq.~(\ref{vwcon}) are {\it not} sufficient to ensure a local maximum.  Examples will be given in Sec.~III~C below.

Proposition 2 applies to observables having an arbitrary number of possible outcomes.  However, it is also of interest to consider the case where $A$ and $B$ are restricted to have a maximum of $n$ possible outcomes, i.e., where the corresponding POMs have at most $n$ non-zero elements. The completeness constraints in Eq.~(\ref{complete}) imply that $n\geq d_1, d_2$.  The maximum of the coincidence rate over such observables, for a given density operator $\rho$, will be denoted by $C^{(n)}_{max}(\rho)$.  Noting the above Lemma, one has
\begin{equation} \label{cn}
C^{(n)}_{max}(\rho) = \max_{X_n,Y_n} C(X_n,Y_n|\rho) = \max_{X_n,Y_n} 
\sum_{j,k=1}^n  \langle x_j,y_j|\rho|x_j,y_j\rangle ,
\end{equation}
where the maximum is over all maximal orthogonal POMs $X_n$ and $Y_n$ on $H_n$.  Clearly, $C^{(n)}_{max}(\rho)$ is a non-decreasing function of $n$, and converges to $C_{max}(\rho)$, i.e., defining $d:=\max \{d_1,d_2\}$,
\begin{equation} \label{chain} 
C^{(d)}_{max}(\rho)  \leq C^{(n)}_{max}(\rho) \leq C^{(\infty)}_{max}(\rho) =C_{max}(\rho). 
\end{equation} 
An explicit expression for $C^{(2)}_{max}(\rho)$ is given in Sec.~IV, and general upper bounds for $C^{(n)}_{max}(\rho)$ are obtained in Sec.~V.  

Now, a maximal POM $A$ with $n$ elements may trivially be extended to an infinite number of elements by defining $|a_j\rangle:=0$ for $j>n$.  Hence, such $n$-valued POMs may be thought of as lying on the `boundary' of the set of {\it all} maximal POMs.  It would be of interest to show that $C^{(n)}_{max}(\rho)$, corresponding to the maximum of coincidence rate over a restricted portion of this boundary, is also (at the least) a {\it local} maximum of coincidence rate with respect to the {\it full set} of maximal POMs.  The following corollary to Proposition 2 shows that the conditions in Eq.~(\ref{vwcon}) are sufficient for this to be the case.

{\bf Corollary}:  {\it If the Hermitian operators $V$ and $W$ defined in Eq.~(\ref{vw}) satisfy $V \geq \langle b_j|\rho|b_j\rangle$ and $W \geq \langle a_j|\rho|a_j\rangle$ for all $j$, for maximal POMs $A^{(n)}$ and $B^{(n)}$ achieving $C^{(n)}_{max}(\rho)$, then $C^{(n)}_{max}(\rho)$ is a local maximum of coincidence rate with respect to the set of} all {\it maximal POMs}.

\begin{proof}
By the above Lemma, maximal POMs with at most $n$ non-zero elements can be represented by the set of maximal orthogonal POMs on $H_n$.  Further, since the group of unitary transformations $U(n)\times U(n)$ is compact, the global maximum of coincidence rate over such orthogonal POMs must be actually be achievable, by two orthogonal POMs $X_n$ and $Y_n$ on $H_n$, having eigenstates $|x_1\rangle,\dots,|x_n\rangle$ and $|y_1\rangle,\dots,|y_n\rangle$ respectively.  It may be shown, just as per the proof of Proposition 2, that these eigenstates must satisfy Eqs.~(\ref{prop2a}) and (\ref{prop2b}) with the ranges of $j,k,l$ restricted $1,2,\dots,n$ (and with $M$ and $N$ restricted to $H_n$). Further, any extension of $X_n$ and $Y_n$ to orthogonal POMs $X$ and $Y$ on $H_\infty$ must satisfy $E|x_j\rangle = 0 = F|x_j\rangle$ for all $j>n$.  It follows for such $X$ and $Y$ that (i) Eq.~(\ref{prop2a}) is trivially satisfied (implying the corresponding POMs $A$ and $B$ are extremal); (ii) the first and second terms of Eq.~(\ref{prop2b}) are the same as for $X_n$ and $Y_n$ when $j\leq n$, and nonnegative when $j>n$ (as a consequence of the premise of the Corollary); and (iii) the third term in Eq.~(\ref{prop2b}) is the same as for $X_n$ and $Y_n$ when $j\leq n$, and vanishes when $j>n$. Hence, from Proposition 2, $C^{(n)}_{max}(\rho)$ is a local maximum of coincidence rate with respect to POMs having an {\it arbitrary} number of elements. 
\end{proof}

Note from the above proof that the maximal POMs $A^{(n)}$ and $B^{(n)}$ achieving $C^{(n)}_{max}(\rho)$ must satisfy Eq.~(\ref{prop1a}), with $k$ and $l$ restricted to the range $1,2,\dots,n$.  It may be checked (noting that $\rho$ is Hermitian) that this places $2n(n-1)$ real constraints on the elements of $A^{(n)}$ and $B^{(n)}$, which are invariant under the $n!$ permutations of the elements that preserve the condition $k\neq l$.  On the other hand, to specify two arbitrary maximal POMs, each having no more than $n$ non-zero elements, requires $2n(n-1)$ real parameters (corresponding to specifying the unitary transformations $|x_j\rangle=U_X|z_j\rangle$, $|y_j\rangle=U_Y|z_j\rangle$ on $H_n$ relative to some fixed orthonormal basis $\{|z\rangle\}$, up to arbitary phases), with $(n!)^2$ possible orderings of the elements (i.e., $n!$ orderings for each POM).  It is therefore expected, for a generic density operator $\rho$, that there are $n!$ pairs of extremal candidates for $A^{(n)}$ and $B^{(n)}$ (for density operators having particular symmetries, there will be further extrema, as per the last paragraph of Sec.~III~A).  However, it is conjectured in the next subsection that $C^{(n)}_{max}(\rho)$ is in fact {\it independent} of $n$, which is equivalent to equality throughout in Eq.~(\ref{chain}).  If true, this means that no more than $d!$ candidates for the optimal observables need be checked in the generic case.

\subsection{Two examples and one conjecture}

As a first example, we will consider the case of `trine' measurements on a two-qubit system, for which the measurement on each qubit optimally distinguishes between the states of the ensemble prepared by the measurement on the other qubit, and vice versa.  Surprisingly, these measurements do {\it not} generate a global (or even a local) maximum of coincidence rate.  

In particular, let $\{|1\rangle, |2\rangle\}$ be a basis set for either qubit, and consider the 3-valued `trine' observables $A\equiv B\equiv\{\frac{2}{3}|\phi_j\rangle\langle\phi_j|\}$, where the normalised kets
\[  |\phi_1\rangle:=|1\rangle ,~~~|\phi_2\rangle:=\frac{1}{2}\left( |1\rangle + \sqrt{3}\,|2\rangle \right) ,~~~|\phi_2\rangle:=\frac{1}{2}\left( |1\rangle - \sqrt{3}\,|2\rangle \right)  \]
form the vertices of an equilateral triangle in the Bloch representation.  For the pure bipartite state $\rho=|\psi\rangle\langle\psi|$, with 
\[ |\psi\rangle:=\frac{1}{\sqrt{2}} \left( |1\rangle\otimes|1\rangle + |2\rangle\otimes|2\rangle \right) ,\]
it is then easily checked that 
\[  \langle a_j|\rho|a_j\rangle = \frac{1}{3}|\phi_j\rangle\langle \phi_j| = \langle b_j|\rho|b_j\rangle  \]
on the respective components. It follows that the operators $V$ and $W$ defined in Eq.~(\ref{vw}) are each equal to $\frac{1}{3} \hat{1}$, implying from Proposition 1 that $A$ and $B$ generate an extremal value of coincidence rate, given by 
\[ C(A,B|\rho) = {\rm tr}[V] = 2/3. \]
Further, the conditions in Eqs.~(\ref{vwcon}) are trivially satisfied for this example, implying via Eq.~(\ref{helstrom}) that $A$ optimally distinguishes between members of the ensemble of states $\{ |\phi_j\rangle\langle \phi_j|; \frac{1}{3}\}$ prepared by measurement of $B$, and vice versa (see also Sec.~IV.1(a) of Ref.~\cite{helstrom}).

However, $A$ and $B$ above do {\it not} generate a global maximum of coincidence rate for state $\rho$, as the maximum possible value of unity may be achieved by instead choosing POMs with elements diagonal with respect to any Schmidt decomposition of $|\psi\rangle$.  Indeed, $A$ and $B$ above do not even generate a {\it local} maximum of coincidence rate - the extremal value of $2/3$ in fact corresponds to a saddle point.  To see this, note first that the optimal distinguishing property implies that varying either $A$ or $B$ (while keeping the other fixed) must {\it decrease} the coincidence rate.  Hence, the extremal value of $2/3$ represents a {\it maximum} with respect to such variations.  On the other hand, consider the one-parameter `mirror-symmetric' family of observables $A^{(\alpha)}\equiv B^{(\alpha)} \equiv \{f_j{(\alpha)}|\phi_j^{(\alpha)}\rangle\langle\phi_j^{(\alpha)}|\}$, with $0\leq\alpha\leq 1$, $f_1{(\alpha)}=1-\alpha$, $f_{2,3}{(\alpha)}=(1+\alpha)/2$, and \cite{mirror}
\[ |\phi_1^{(\alpha)}\rangle := |1\rangle,~~~~|\phi_{2,3}^{(\alpha)}\rangle = (1+\alpha)^{-1/2}\left(\sqrt{\alpha} |1\rangle \pm  |2\rangle \right) . \]
Choosing $\alpha=1/3$ corresponds to the trine observables.  It is straightforward to calculate
\[  C(A^{(\alpha)},B^{(\alpha)}|\rho) = \frac{2}{3} + \frac{3}{4}\left(\alpha-\frac{1}{3}\right)^2 , \]
and hence the extremal value of $2/3$ represents a {\it minimum} of coincidence rate with respect to the variation of $\alpha$ \cite{footnote}.

As an example of Proposition 2, consider now a separable state of the form
\begin{equation} \label{rhosep}
\rho=\sum_{j=1}^d \lambda_j\,|\psi_j\rangle\langle\psi_j|\otimes |\chi_j\rangle\langle \chi_j| , 
\end{equation}
where the mutual orthogonality property $\langle\psi_j|\psi_k\rangle = \delta_{jk}$ 
is satisfied, and each $|\chi_j\rangle$ is arbitrary.  Let $A$ be the maximal orthogonal POM defined by $|a_j\rangle :=|\psi_j\rangle$, i.e., the optimal POM for distinguishing members of the ensemble $\{|\psi_j\rangle\langle\psi_j|;\lambda_j\}$; and let $B$ be the maximal POM which optimally distinguishes between members of the pure-state ensemble $\{|\chi_j\rangle\langle\chi_j|;\lambda_j\}$ (the existence of such a maximal POM $B$ follows from Theorem 2 of Ref.~\cite{eldar}).  Thus, from Eqs.~(\ref{vw}) and (\ref{helstrom}), 
\[ \left(V -\lambda_j|\psi_j\rangle\langle \psi_j|\right)|a_j\rangle = 0 =\left(W -\lambda_j|\chi_j\rangle\langle \chi_j|\right)|b_j\rangle,~~~~V\geq \lambda_j |\psi_j\rangle\langle \psi_j| ,~~ W\geq \lambda_j |\chi_j\rangle\langle \chi_j| . \]
It is then straightforward to check that both conditions of Proposition 2 are satisfied by any $X$ and $Y$ corresponding to $A$ and $B$ respectively (in particular, the third term in Eq.~(\ref{prop2b}) vanishes identically, since the orthogonality of the elements of $A$ implies that $\rho$ and $|x_j\rangle\langle x_j|$ must commute).  Hence this choice of $A$ and $B$ generates a local maximum of coincidence rate.
Indeed, since a measurement outcome $A=a_j$ for the first component is perfectly correlated with preparation of state $|\chi_j\rangle$ for the second component, and since $B$ is the best possible measurement for distinguishing between such prepared states, the above choice of $A$ and $B$ is intuitively expected to generate a {\it global} maximum of coincidence rate.   

Note that if $d=2$ in Eq.~(\ref{rhosep}), then $B$ is the orthogonal POM generated by the eigenstates of \cite{helstrom}
\[ \eta:=\lambda_1|\chi_1\rangle\langle\chi_1| - \lambda_2|\chi_2\rangle\langle\chi_2|. \]
The corresponding maximum value of coincidence rate follows as (cf. Eq.~(2.34) in Chap.~IV of Ref.~\cite{helstrom})
\begin{equation} \label{c2}
 C(A,B|\rho)=\frac{1}{2}\left(1+{\rm tr}[|\eta|]\right) = \frac{1}{2}\left[ 1+ \left( 1- 4\lambda_1\lambda_2 |\langle\chi_1|\chi_2\rangle|^2\right)^{1/2}  \right] .
\end{equation}
This result is significantly generalised in Sec.~IV.

Finally, note that in the above example that $A$ is an {\it orthogonal} POM, having the minimum possible number, $n=d$, of non-zero elements.  We conjecture that this may be an instance of a general rule.  As motivation, observe that if Alice and Bob each measure observables having $n\geq d$ possible outcomes, then the outcomes will typically have a greater degree of randomness when $n>d$. For example, the entropy $H(A)$ of a maximal POM $A$ is bounded below by
\[   H(A) = -\sum_j p_j \log_2 p_j \geq -\log_2 \max_j p_j \geq  - \log_2  \max_j \langle a_j|a_j\rangle , \]
which is nontrivial if $A$ is non-orthogonal (i.e., if $n>d$). Similarly, the joint entropy of $A$ and $B$ is bounded below by
 \[  H(AB) \geq - \log_2 \max_{j,k} \langle a_j|a_j\rangle\,\langle b_k|b_k\rangle. \]
Further, the more random a distribution is, the more spread out it is over the set of possible outcomes. Hence, the sum over the diagonal elements of the joint distribution $p_{jk}$ (i.e., the coincidence rate), will typically be smaller.  It follows that choosing $n>d$ is typically expected to have a decreasing effect on the maximum achievable coincidence rate:

{\bf Conjecture}: {\it The global maximum of coincidence rate, for a bipartite density operator with finite support, can always be achieved by  observables $A$ and $B$ having at most $d=\max \{d_1,d_2\}$ possible outcomes, where $d_1$ and $d_2$ are the Hilbert space dimensions defined in Eq.~(\ref{dim})}.

Note that the conjecture implies at least one of $A$ and $B$ corresponds to an orthogonal POM, depending on whether $d=d_1$ and/or $d=d_2$. Note further that the conjecture corresponds to the case of equality throughout in Eq.~(\ref{chain}), i.e., to the condition
\begin{equation} \label{conj}
C_{max}(\rho) \equiv C^{(d)}_{max}(\rho) . 
\end{equation}
This conjecture is consistent with the convexity properties discussed in Sec.~II and, if true, would greatly simplify the numerical determination of the maximum coincidence rate, as only POMs with $d$ elements would need to be considered.  
Partial numerical support has been found for the conjecture, for the case of two-qubit systems.  In particular, the evaluation of coincidence rate for $\approx 10^{11}$ pairs of maximal POMs having no more than 3 non-zero elements, for each member of a random sample of $1 200$ bipartite density operators, indicates that $C^{(2)}_{max}\equiv C^{(3)}_{max}$.  

\section{Maximum spin correlation for two qubits}

An exact result for two-qubit systems is derived here, which also introduces the basic method used in the following section to derive general upper bounds for the coincidence rate.

A system of two qubits is described by a density operator $\rho$ on $H_2\otimes H_2$, so that $d_1=d_2=d=2$.  Consider the problem of finding the maximal {\it two}-valued POMs $A$ and $B$ which maximise the coincidence rate.  Such POMs are necessarily orthogonal, corresponding to the measurement of spin in some direction, and hence, noting Eq.~(\ref{cn}), the corresponding coincidence rate can be written as
\begin{equation} \label{cspin}  
C^{(2)}_{max}(\rho) = \max_{a,b} C(\sigma^{(1)}\cdot a,\sigma^{(2)}\cdot b|\rho) ,
\end{equation}
where $a$ and $b$ are unit directions.  Note that $C^{(2)}_{max}(\rho)$ is in fact equal to the global maximum of coincidence rate,  $C_{max}(\rho)$, if the conjecture in Eq.~(\ref{conj}) is correct.

To determine $C^{(2)}_{max}(\rho)$, let $|m\rangle$ denote the $+1$ eigenstate of $\sigma\cdot m$ for unit direction $m$, so that $|m\rangle\langle m|=(1+\sigma\cdot m)/2$.  Hence, $A\equiv \{ |a\rangle\langle a|, |-a\rangle\langle -a|\}$, $B\equiv \{ |b\rangle\langle b|, |-b\rangle\langle -b|\}$, and the coincidence rate follows via Eq.~(\ref{coinc}) as
\[ C(A,B|\rho) = \frac{1}{2} {\rm tr}[\rho (1 + \sigma^{(1)}\cdot a\otimes \sigma^{(2)}\cdot b)] = \frac{1}{2} ( 1 + a^T S b ) ,\]
where $S$ is the $3\times 3$ `spin correlation' matrix defined by
\begin{equation} \label{sjk}
S_{jk} := \langle \sigma^{(1)}_j \otimes \sigma^{(2)}_k \rangle .
\end{equation}
Note that $S$ is real, but in general is not symmetric.

Now, the singular value decomposition theorem \cite{horn} states that any real $p\times q$ matrix $S$ can be put in the form
\begin{equation} \label{svd} 
S = R_1DR_2^T ,
\end{equation}
where $R_1$ and $R_2$ are real orthogonal matrices (of dimensions $p\times p$ and 
$q\times q$ respectively), and $D$ is a real $p\times q$ matrix of the form
\[ D_{jk} = s_j \delta_{jk},~~~~~s_1\geq s_2\geq \dots \geq 0. \]
The numbers $s_j$ are called the singular values of $S$, and are just the square roots of the eigenvalues of each of $S^T S$ and $S
S^T$, while $R_1$ and $R_2$ are formed by the respective eigenvectors of $S^T S $ and $SS^T$ \cite{horn}. The largest singular value, $s_1$, is also known as the {\it spectral norm} of $S$.

It follows in particular, defining $u=R_1^Ta$ and $v=R_2b$, and using the Schwarz inequality, that for unit vectors $a$ and $b$ one has
\begin{eqnarray*}
\max_{a,b} a^TSb &=& \max_{u,v} |u^T D v| \\
&=& \max_{u,v} \left|\sum_j (\sqrt{s_j}u_j)(\sqrt{s_j}v_j)\right|\\ &\leq& \max_{u,v} [ \sum_j s_j(u_j)^2]^{1/2}~ [ \sum_k s_k 
(v_k)^2 ]^{1/2}\\
&=& \max_{u} \sum_j s_j (u_j)^2 \leq (\max_j s_j) \sum_j (u_j)^2 = s_1 ,
\end{eqnarray*}
with equality obtained for the choice $u=v=x:=(1,0,0)$.  Thus, 
\begin{equation} \label{chmax} 
C^{(2)}_{max}(\rho) = \frac{1}{2}(1+s_1) , 
\end{equation}
where $s_1$ is the spectral norm of the spin-correlation matrix $S$ defined in Eq.~(\ref{sjk}) (hence one must have $s_1=1$ for all pure states), with this maximum coincidence rate being achieved via spin measurements in the directions 
\[ a=R_1x,~~~~~~b=R_2x .  \]
The case of spin measurements on two qubits is thus completely solved. 

As a simple example, consider the separable state
\[ \rho = \lambda_1 |z\rangle\langle z|\otimes \tau_1 + \lambda_2 |-z\rangle\langle -z|\otimes \tau_2 \]
for arbitary qubit density operators $\tau_1$ and $\tau_2$.  One finds that all elements of the spin-correlation matrix vanish other than the third row, which is given by the 3-vector $r$ with components
\[ r_k:=\lambda_1{\rm tr}[\tau_1 \sigma^{(2)}_k] - \lambda_2{\rm tr}[\tau_2 \sigma^{(2)}_k] . \]
It follows that only the $33$-component of $SS^T$ is non-zero, and equal to $r\cdot r$, yielding 
\[ C^{(2)}_{max}(\rho) = (1+|r|)/2.  \]
This result generalises Eq.~(\ref{c2}) of the previous section, and greatly simplifies calculation of the corresponding coincidence rate, as it does not require explicit diagonalisation of the operator $\eta$.  Note that the coincidence rate is equal to the average probability for optimally discriminating between members of the ensemble $\{ \tau_j; \lambda_j\}$ \cite{helstrom}.

As a second example, consider the isotropic state \cite{werner}
\[ \rho_w = w|\Psi^-\rangle\langle\Psi^-| + \frac{1-w}{3}\hat{1}_T,  \]
where $|\Psi^-\rangle$ denotes the singlet state, $\hat{1}_T=\hat{1}-|\Psi^-\rangle\langle\Psi^-|$ denotes the unit operator on the triplet subspace, and $0\leq w\leq 1$.  This state is rotationally-invariant, and the spin-correlation matrix is easily calculated to be $S_{jk}=-(4w-1)\delta_{jk}/3$. It follows immediately that
\[ C^{(2)}_{max}(\rho_w) = \frac{1}{2} \left(1 + \frac{|4w-1|}{3}\right) , \]
with the maximum coincidence rate being achieved by the choice $a=b$ for $0\leq w\leq 1/4$, and $a=-b$ for $1/4\leq w\leq 1$.

Finally, for a general factorisable state $\rho=\rho_1\otimes\rho_2$, with $\rho_1=(1+ m\cdot\sigma)/2$ and $\rho_2=(1+ n\cdot\sigma)/2$, the spin correlation matrix is just the outer product $S = m\,n^T $, so that $SS^T = (n\cdot n) mm^T$,
with eigenvalues $(m\cdot m)(n\cdot n)$, 0, and 0.  It follows immediately from Eq.~(\ref{chmax}) that the maximum possible coincidence rate for two uncorrelated qubits is given by
\[ C^{(2)}_{max}(\rho_1\otimes\rho_2) = \frac{1}{2} \left( 1 + |m|\,|n|\right) , \]
achieved by the choice of the measurement directions $a=m$ and $b=n$.

\section{Bounds for coincidence rate}  

\subsection{A general upper bound}

Here an upper bound is given for $C^{(n)}_{max}(\rho)$ in Eq.~(\ref{cn}), i.e., for the maximum achievable coincidence rate when Alice and Bob are restricted to measurements of $n$-valued observables.  This bound is tight for the case $n=2$, reducing to Eq.~(\ref{chmax}) above.  Conversely, taking the limit $n\rightarrow \infty$ gives a global upper bound for $C_{max}(\rho)$, which turns out to be equal to the computable cross norm of $\rho$ \cite{ccn}.  Note that if the conjecture in Eq.~(\ref{conj}) is correct, then taking $n=d$ will give a much tighter bound in general for $C_{max}(\rho)$.

First, it is well known that the traceless Hermitian operators on an $n$-dimensional Hilbert space $H_n$ form a real vector space of dimension $n^2-1$, with inner product $(M,N):={\rm tr}[MN]$ \cite{bloch}.  Hence, if $\{K_p\}$ and $\{ L_q\}$ denote two orthonormal basis sets for this vector space, then 
\begin{equation} \label{orthog} 
{\rm tr}[K_pK_{q}] = \delta_{ij} = {\rm tr}[L_pL_{q}] ,~~~~~K_p = \sum_q R_{pq} L_q , 
\end{equation}
for some orthogonal matrix $R$ (i.e., $RR^T=I$). It follows that the trace-free part of any operator $Z$ on $H_n$ can be written as
\begin{equation} \label{zrep}
Z - \frac{{\rm tr}[Z]}{n}\hat{1} = \sum_p {\rm tr}[ZK_p]\,K_p =  \sum_q {\rm tr}[ZL_q]\,L_q ,\
\end{equation}
and that any bipartite density operator $\rho$ on $H_n\otimes H_n$ can be expressed as 
\begin{equation} \label{rep}
\rho = \frac{1}{n^2} \hat{1}\otimes \hat{1}+ \sum_{p} u_p \,K_p\otimes \hat{1} + \sum_{q} v_q \, \hat{1}\otimes L_q + \sum_{p,q} T_{pq}\,K_p\otimes L_q ,
\end{equation}
where
\begin{equation} \label{t}
u_p:=\langle K_p\otimes \hat{1}\rangle/n,~~~v_q:=\langle \hat{1}\otimes L_q\rangle/n,~~~T_{pq}:=\langle K_p\otimes L_q\rangle .
\end{equation}
This is referred to as a Fano form for $\rho$ \cite{fano, fanoz}.  

Now, using Eqs.~(\ref{orthog})-(\ref{t}), the coincidence rate  for two maximal orthogonal POMs $X_n\equiv\{|x_j\rangle\langle x_j|\}$ and $Y_n\equiv\{|y_j\rangle\langle y_j|\}$ on $H_n$ simplifies to
\[ C(X_n,Y_n|\rho) =  1/n + {\rm Tr}[{T}RW], \]
where ${\rm Tr}$ denotes the {\it matrix} trace, and 
\begin{equation} \label{w} 
W_{pq}:=\sum_j \langle x_j|L_p|x_j\rangle \, \langle y_j|L_q|y_j\rangle .
\end{equation}
Further, Eq.~(7.4.14) of Ref.~\cite{horn} implies, for any two real matrices $T$ and $W$ and orthogonal matrix $R$, that
\[ \left| {\rm Tr}[TRW]\right| \leq \sum_k s_k(T) s_k(W) , \]
where $s_1(P)\geq s_2(P)\geq\dots$ denote the singular values of matrix $P$ (see Sec.~IV).  Hence, noting Eq.~(\ref{cn}), one has
\begin{equation} \label{horn}
C^{(n)}_{max}(\rho) \leq 1/n + \sum_k s_k(T) s_k(W) . 
\end{equation}

To simplify this upper bound, note first that $W$ can be written, 
in terms of the vectors 
\[ {\bm f}^{(j)}_p := \langle x_j|L_p|x_j\rangle,~~~~{\bm g}^{(j)}_p := \langle y_j|L_p|y_j\rangle , \]
as the sum of outer products $W = \sum_j {\bm f}^{(j)}\,({\bm g}^{(j)})^T$.
Using Eq.~(\ref{zrep}) one finds
\[  {\bm f}^{(j)}\cdot {\bm f}^{(k)} = \delta_{jk} - 1/n = {\bm g}^{(j)} \cdot {\bm g}^{(k)} , \]
implying that 
\[ W^TW=\sum_j {\bm g}^{(j)}\,({\bm g}^{(j)})^T =(W^TW)^2,~~~~{\rm Tr}[W^TW]=n-1. \]  
Thus, $W^TW$ is an $(n-1)$-dimensional projection matrix, implying that the non-zero singular values of $W$ consist of precisely $n-1$ $1$s.  The above upper bound therefore reduces to
\begin{equation} \label{inter} C^{(n)}_{max}(\rho) \leq 1/n + \sum_{k=1}^{n-1} s_k({T}) , \end{equation} 
where the matrix $T$ is defined in Eq.~(\ref{t}).

The bound can be further simplified, via a judicious choice of the basis sets $\{K_p\}$ and $\{ L_q\}$.  In particular, recall that $\rho$ only has support on the subspace $H_1\otimes H_2$ of $H_n\otimes H_n$ (see Sec.~III~B).  The first $(d_1)^2-1$ elements of $\{K_1, K_2,\dots\}$ can therefore be chosen to form a basis set for the traceless operators on $H_1$, and the first $(d_2)^2-1$ elements of $\{L_1,L_2,\dots\}$ can similarly be chosen to form a basis set for the traceless operators on $H_2$.  Two further basis elements, relabelled as $K_0$ and $L_0$ for convenience, will be chosen to have the forms
\[ K_0 := \alpha_1 E - \beta_1 (\hat{1} -E),~~~~L_0 := \alpha_2 F - \beta_2 (\hat{1} -F), \]
where $E$ and $F$ denote the projections from $H_n$ to $H_1$ and $H_2$. The requirements ${\rm tr}[K_0] = {\rm tr}[L_0]=0$ and ${\rm tr}[(K_0)^2] = {\rm tr}[(L_0)^2]=1$ imply that
\[ \alpha_1=\left( 1/d_1-1/n\right)^{1/2},~~~\beta_1 = \alpha_1d_1/(n-d_1) , \]
\[ \alpha_2=\left( 1/d_2-1/n\right)^{1/2},~~~\beta_2 = \alpha_2d_2/(n-d_2) . \]
Since the remaining basis elements must be orthogonal to the above basis elements, they cannot contribute to the Fano form of $\rho$ in Eq.~(\ref{rep}).  Hence, using Eq.~(\ref{t}), the only nonzero rows and columns of the matrix $T$ are given by the $(d_1)^2\times (d_2)^2$-submatrix
\begin{equation} \label{tprime}  
\ T^{(n)} := \left( \begin{array}{cc}
\langle K_0\otimes L_0\rangle & \langle K_0\otimes L_q\rangle \\
\langle K_p\otimes L_0\rangle & \langle K_p\otimes L_q\rangle
\end{array} \right)
= \left( \begin{array}{cc}
\alpha_1\alpha_2 & \alpha_1 \langle \hat{1}_1\otimes L_q\rangle  \\
\alpha_2 \langle K_p\otimes\hat{1}_2\rangle  & \langle K_p\otimes L_q\rangle
\end{array} 
\right) ,
\end{equation}
where $\hat{1}_1$ and $\hat{1}_2$ denote the identity operators on $H_1$ and $H_2$ respectively, and $1\leq p\leq (d_1)^2-1$, $1\leq q\leq (d_2)^2-1$.  Substitution into Eq.~(\ref{inter}) yields the main result of this section:

{\bf Theorem}: {\it The maximum coincidence rate obtainable for a bipartite state with finite support, via maximal POMs having no more than $n$ nonzero elements, is bounded above by}
\begin{equation} \label{theo}
C^{(n)}_{max}(\rho) \leq 1/n + \sum_{k=1}^{\min\{ n-1, \delta^2\} } s_k({T^{(n)}})
\end{equation}
{ \it where $\delta:= \min\{d_1,d_2\}$, and the matrix $T^{(n)}$ is defined in Eq.~(\ref{tprime})}.  

Since local unitary transformations correspond to left and right multiplication of $T^{(n)}$ by orthogonal matrices, which leave the singular values unchanged \cite{horn}, this upper bound is invariant under such transformations.  

For the case of two qubits, with $n=d_1=d_2=2$, one may choose $K_p=\sigma^{(1)}_p/\sqrt{2}$ and $L_q=\sigma^{(1)}_q/\sqrt{2}$. The `zeroth' row and column of $T^{(n)}$ vanish for this case, since $\alpha_1=\alpha_2=0$, leaving a $3\times 3$-submatrix equal to one-half of the spin-correlation matrix $S$ in Eq.~(\ref{sjk}).  Thus, for this case, the upper bound of the theorem reduces to $(1+s_1(S))/2$, which can in fact always be achieved, as per Eq.~(\ref{chmax}) of the previous section.
However, for $n>3$ the upper bound in Eq.~(\ref{theo}) cannot always be attained, essentially because the set of orthogonal matrices $R$ in Eq.~(\ref{orthog}) is larger than the set of unitary transformations on $H_n$ \cite{bloch}. 

\subsection{Examples}

Note first that taking the limit $n\rightarrow\infty$ in Eq.~(\ref{theo}) yields a {\it global} upper bound for the coincidence rate, independent of the possible number of measurement outcomes:
\begin{equation}  \label{cgmax}
C_{max}(\rho) \leq \sum_{k=1}^{\delta^2} s_k(T^{(\infty)}) = {\rm Tr}\left[\sqrt{(T^{(\infty)})^TT^{(\infty)}}\right] = \left\| T^{(\infty)} \right\|_{\rm Tr} .
\end{equation}
Thus, the upper bound is just the trace norm of $T^{(\infty)}$.  Noting that $\alpha_1\rightarrow 1/\sqrt{d_1}$ and $\alpha_2\rightarrow 1/\sqrt{d_2}$ in this limit, it follows that the coefficients of $T^{(\infty)}$ yield a Fano form for $\rho$ on $H_1\otimes H_2$, via
\begin{equation} \label{frep}
\rho = T^{(\infty)}_{00} \hat{1}_1\otimes \hat{1}_2/\sqrt{d_1d_2}+ \sum_{p\geq 1} T^{(\infty)}_{p0} \,K_p\otimes \hat{1}_2/\sqrt{d_2} + \sum_{q\geq 1} T^{(\infty)}_{0q} \, \hat{1}_1\otimes L_q/\sqrt{d_1} + \sum_{p,q\geq 1} T^{(\infty)}_{pq}\,K_p\otimes L_q .
\end{equation}
The trace norm of $T^{(\infty)}$ may therefore be recognised as the `computable cross norm' measure of quantum entanglement \cite{fanoz,ccn}, i.e., {\it the maximum possible coincidence rate, $C_{max}(\rho)$, is bounded above by the computable cross norm}.  

The computable cross norm cannot be greater than unity for any separable states \cite{fanoz,ccn}, and hence the upper bound in Eq.~(\ref{cgmax}) is always nontrivial for separable states (and for a large proportion of entangled states).  However, a stronger bound is postulated further below.

Second, it is of interest to consider measurements restricted to the {\it minimum} number of possible measurement outcomes, i.e., with $n=d$. For this case $\alpha_1=\alpha_2=0$, implying that the only nonvanishing part of $T^{(d)}$ is the $(d_1^2-1)\times(d_2^2-1)$ submatrix $\tilde{T}_{pq}=\langle K_p\otimes L_q\rangle$ with $p,q\geq1$,  yielding 
\begin{equation} \label{cdmax}
C^{(d)}_{max}(\rho) \leq 1/d + \sum_{k=1}^{\min\{ d-1, \delta^2-1\} } s_k({\tilde{T}}) . 
\end{equation}
As noted above, this bound is tight for the case $d_1=d_2=2$. For two-qudit systems it bounds the coincidence rate for the case of measurements described by orthogonal POMs.   It may also be noted, in analogy to the computable cross norm above, that $\tilde{T}$ has similarly been used in partial characterisations of entanglement \cite{vicente,gunnar}.   For example, the trace norm of $\tilde{T}$ is never greater than $ [(1-1/d_1)(1-1/d_2)]^{1/2}$ for any separable state \cite{vicente}. These general underlying connections, between bounds for correlations and measures of entanglement, would be an interesting subject for further investigation (see also Sec.~VI).

Third, a simple yet general example of the Theorem is provided by the Werner state for two qudits \cite{werner}, which has the Fano form \cite{vicente}
\[ \rho_x :=  \frac{1}{d^2} \hat{1}\otimes \hat{1} + \frac{x-1/d}{d^2-1} \sum_{p\geq 1} K_p\otimes K_p , \]
with $-1\leq x\leq 1$.  It follows via Eqs.~(\ref{rep}), (\ref{t}) and (\ref{tprime}) that $T^{(n)}$ is diagonal, and so, noting that $\alpha_1\alpha_2=1/d-1/n$ for this case, the Theorem yields
\[ C^{(n)}_{max}(\rho_x) \leq  \frac{1}{n} + (D-1)\frac{|x-1/d|}{d^2-1} + \max\left\{\frac{1}{d}-\frac{1}{n}, \frac{|x-1/d|}{d^2-1}\right\} ,\]
where $D:=\min\{n-1, d^2\}$.  Note that, in the limit $n\rightarrow\infty$, the righthand side approaches the computable cross norm for Werner states, $1/d+|x-1/d|$, as expected \cite{ccn}.  It is also straightforward to verify via direct calculation that this bound is {\it tight} for the case $n=d$ and $x\geq1/d$, i.e.,
\begin{equation}
C^{(d)}_{max}(\rho_x) = 1/d + |x-1/d|/(d+1) 
\end{equation}
for $x\geq 1/d$, achieved by the choice $A=B$.  For $x<1/d$, a modification of the Theorem for negative definite $\tilde{T}$ gives the tight upper bound $C^{(d)}_{max}(\rho_x) = 1/d + |x-1/d|/(d^2-1)$, achieved by maximal orthogonal POMs satisfying  $|a_j\rangle =|b_{P(j)}\rangle$ for any permutation $P$ of $1,2,\dots,d$ with $P(j)\neq j$ for all $j$.  This example may be regarded as a generalisation of the $d=2$ isotropic example in Sec.~IV, where one identifies $x$ with $1-2w$.

Note finally that if the conjecture in Eq.~(\ref{conj}) is correct, then the bound in Eq.~(\ref{cdmax}) is in fact an upper bound for $C_{max}(\rho)$, which is generally much tighter than the computable cross norm bound in Eq.~(\ref{cgmax}).  
For example, consider any state for which the reduced density operators are maximally random, i.e., where $\rho_1=\hat{1}_1/d_1$ and $\rho_2=\hat{1}_2/d_2$ (eg, the Werner state $\rho_x$ considered above).  It then follows trivially via Eq.~(\ref{tprime}) that 
\[ \left\| T^{(\infty)} \right\|_{\rm Tr} = (d_1d_2)^{-1/2} + \left\| \tilde{T} \right\|_{\rm Tr} \geq 1/d + \left\| \tilde{T} \right\|_{\rm Tr} . \]
Thus, for such states, the bound in Eq.~(\ref{cdmax}) is never greater than that in Eq.~(\ref{cgmax}), and is generally smaller whenever $d\leq \delta^2$.  

\section{Conclusions} 

It is well known that determining the maximum mutual information between the components of a given bipartite system is a difficult problem \cite{davies}.  The results of this paper indicate that it is similarly not a straightforward matter to maximise the coincidence rate, despite (i) its linearity with respect to the density operator, and (ii) formal similarities with the well known problem of optimal state discrimination. A notable exception is the case of spin measurements on two-qubit systems, which has been fully solved in Sec.~IV.  More generally, one only has available the formal equations for the correlation basis derived in Propositions 1 and 2 of Sec.~III, and the upper bounds for $n$-valued measurements derived in the Theorem of Sec.~IV.  These general results could be substantially strengthened if the Conjecture of Sec.~III~C could be verified.  It would further be of interest to determine whether or not the `optimal discrimination' conditions in Eq.~(\ref{vwcon}) must be satisfied by observables corresponding to a global maximum of coincidence rate.

It is worth mentioning here some generalisations of the results in Secs.~IV and V, to other linear measures of correlation.  For example, note that the spin correlation matrix $S$ in Eq.~(\ref{sjk}) is closely related to the {\it spin  covariance} matrix $\overline{S}$ defined by
\[  \overline{S}_{jk} := \langle \sigma^{(1)}_j\otimes \sigma^{(2)}_k \rangle - \langle \sigma^{(1)}_j\rangle\,\langle \sigma^{(2)}_k \rangle .\]
In particular, explicitly indicating dependence on the density operator, one has $\overline{S}(\rho) = S(\rho)-S(\rho_1\otimes\rho_2)$.  This covariance matrix has been of recent interest in the characterisation of entanglement \cite{plenio, gunnar}.  For example, the main result in Sec.~IV of Ref.~\cite{gunnar} may be simplified to
\begin{equation}  
{\rm Tr}[\overline{S}^T \overline{S}] = 4\, {\rm tr}[ (\rho-\rho_1\otimes\rho_2)^2] \leq 1  ,
\end{equation}
for all separable two-qubit states, i.e., {\it a separable state $\rho$ can lie at distance of at most $1/2$ from  $\rho_1\otimes\rho_2$, as measured by the Hilbert-Schmidt metric}.

Now, the covariance of two arbitrary spin observables, corresponding to directions $a$ and $b$, may be written as
\[  {\rm Cov}(A, B|\rho) =  a^T \overline{S} b . \]
The methods of Sec.~IV then immediately lead to the upper bound
\begin{equation} \label{cov}
\max_{a,b} {\rm Cov}(A, B|\rho) = s_1(\overline{S})  
\end{equation}
analogous to Eq.~(\ref{chmax}), i.e., {\it the maximum possible spin covariance for state $\rho$ is given by the spectral norm of the spin covariance matrix}.  Note that this bound is invariant under local unitary transformations.  
It follows, for example, that Theorem 1 of Ref.~\cite{plenio} may be strengthened to the observable-independent statement that
\begin{equation} \label{entang}  
{\rm tr}[\rho^2] + \frac{1}{2} s_1(\overline{S}) \leq 1 
\end{equation}
{\it for all separable states of two-qubit systems}.  Noting Eq.~(\ref{cov}), this inequality is also valid if ${\rm Cov}(A, B|\rho)$ is substituted for $s_1(\overline{S})$, for {\it any} spin observables $A$ and $B$. Similarly, Eq.~(24) of Ref.~\cite{plenio} may be strengthened, using the methods of Sec.~IV, to the entanglement bound
\begin{equation}
E_N(\rho) \geq  \max\{ 0, \log_2 [s_1(S) +s_2(S)] \} 
\end{equation}
for the logarithmic negativity of a two-qubit system.  Thus, as in Sec.~V, correlation and entanglement bounds are seen to be closely related.

Finally, consider some general linear measure of correlation, of the form 
\[  G(A,B|\rho) := \sum_{j,k} p_{jk} g_{jk} = \sum_{jk} g_{jk} \langle a_j,b_k|\rho|a_j,b_k\rangle  . \]
Coincidence rate corresponds to the choice $g_{jk}=\delta_{jk}$.  The related `covariance' measure
\[ \overline{G}(A,B|\rho) := G(A,B|\rho) - G(A,B|\rho_1\otimes\rho_2)  \]
then has the desirable property of automatically vanishing for uncorrelated states.  The methods of Sec.~V~A may then be applied to $\overline{G}$, with $\rho$ replaced by $\rho-\rho_1\otimes\rho_2$, to yield the corresponding upper bound
\begin{equation} 
\left|\overline{G}(A^{(n)},B^{(n)}|\rho)\right| = \left| {\rm Tr}[\overline{T}RW]\right| \leq \sum_k s_k(\overline{T})\,s_k(W^G) , 
\end{equation}
analogous to Eq.~(\ref{horn}),  for maximal POMs $A^{(n)}$ and $B^{(n)}$ having $n$ elements each. Here
\[  \overline{T}_{pq}:=\langle K_p\otimes L_q\rangle - \langle K_p\rangle\,\langle L_q\rangle , \]
and the definition of $W$ in Eq.~(\ref{w}) is generalised to 
\[ W^G_{pq}:=\sum_{j,k} g_{jk} \langle x_j|L_p|x_j\rangle \, \langle y_k|L_q|y_k\rangle . \]
This bound is tight for spin measurements on two-qubit systems.  For the choice $g_{jk}=\delta_{jk}$ the bound simplifies to
\begin{equation}
\left| {\rm Corr}(A^{(n)}, B^{(n)}|\rho) \right| \leq \sum_{k=1}^{\min\{n-1,\delta^2-1\} } s_k(\overline{T})
\end{equation}
for the `correlation' $\sum_j(p_{jj}-p_jq_j)$ of any two $n$-valued maximal POMs (see Sec.~II), generalising Eq.~(\ref{cov}) above.  Note that one may simplify the calculation of the above bounds by choosing the basis elements as in Sec.~V, allowing one to replace $\overline{T}$ by the submatrix corresponding to $1\leq p\leq (d_1)^2-1$ and $1\leq q\leq (d_2)^2-1$.

\appendix

\section{Proof of Proposition 2}

To prove Proposition 2 in Sec.~III~B, note first that all infinitesimal variations of the orthogonal POMs $X$ and $Y$ in Eq.~(\ref{cxy}) are generated by infinitesimal unitary transformations on $H_\infty$, and hence are of the form
\[ |x_j\rangle\rightarrow \exp(i\epsilon M)|x_j\rangle,~~~~~|y_j\rangle\rightarrow \exp(i\epsilon N)|y_j\rangle \]
for arbitrary Hermitian operators $M$ and $N$ on $H_\infty$, where $\epsilon$ is a infinitesimal real parameter.  Note from Eq.~(\ref{cxy}) that these variations are equivalent to keeping $X$ and $Y$ fixed and instead varying the density operator, viz.
\[ \rho\rightarrow \rho_\epsilon := \exp(-i\epsilon K)\rho\exp(i\epsilon K), \]
with $K := M\otimes 1 + 1\otimes N$.  Expanding in powers of $\epsilon$ gives
\[ \rho_\epsilon = \rho  -i [K,\rho] - (1/2) \epsilon^2 [K,[K,\rho]] + \dots \]
and hence the corresponding variation in coincidence rate is 
\[ \delta C = -i\epsilon \sum_j \langle x_j,y_j|[K,\rho]|x_j,y_j\rangle -(1/2)\epsilon^2 \sum_j\langle x_j,y_j|[K,[K,\rho]]|x_j,y_j\rangle + \dots . \]
Requiring the first-order variation to vanish yields
\[ 0 = {\rm tr_1}(M\,\sum_j[X_j,\langle y_j|\rho|y_j\rangle ]\,) + {\rm tr_2}(N\,\sum_j [Y_j,\langle x_j|\rho|x_j\rangle]\,) \]
for arbitrary $M$ and $N$, where $X_j$ and $Y_j$ denote $|x_j\rangle\langle x_j|$ and $|y_j\rangle\langle y_j|$ respectively. Hence, each operator sum must vanish identically, and Eq.~(\ref{prop2a}) follows as the matrix components of these sums, with respect to the  $X$ and $Y$ basis sets respectively.  

Requiring the second-order variation to be no greater than zero, as is required for a local maximum, is equivalent to 
\[ \sum_j \left\{ {\rm tr}_1\left([M,[M,X_j]\langle y_j|\rho|y_j\rangle\right) + {\rm tr}_2\left([N,[N,Y_j]\langle x_j|\rho|x_j\rangle\right) + 2\,{\rm tr}\left( \rho\,[M,X_j|]\otimes [N,Y_j|]\right) \right\} \geq 0. \]
Now, defining the Hermitian operators
\[ 
\tilde{V} := \sum_j \langle y_j|\rho|y_j\rangle\,|x_j\rangle\langle x_j| ,~~~~~\tilde{W}:= \sum_j \langle x_j|\rho|x_j\rangle\,|y_j\rangle \langle y_j| ,
\]
and using Eq.~(\ref{aex}) and $\sum_j X_j=1$, the summation over the first term may be simplified to give
\begin{eqnarray*}
\sum_j  {\rm tr}_1\left([M,[M,X_j]\langle y_j|\rho|y_j\rangle\right) &=& \sum_j {\rm tr}_1\left(MX_j\langle y_j|\rho|y_j\rangle M + h.c. -2X_jM \langle y_j|\rho|y_j\rangle M\right)\\
&=& \sum_j {\rm tr}_1 \left( X_jM[\tilde{V}+\tilde{V}^\dagger]M - 2X_jM \langle y_j|\rho|y_j\rangle M\right)\\
&=& 2 \sum_j \langle x_j|M(\tilde{V}-\langle b_j|\rho|b_j\rangle)M|x_j\rangle .
\end{eqnarray*}
The summation over the second term may be similarly simplified in terms of $\tilde{W}$.  Equation~(\ref{prop2b}) then immediately follows if it can be shown that $\tilde{V}=V$ and $\tilde{W}=W$. 

To do so, note first from Eqs.~(\ref{vw}) and (\ref{aex}) that $\tilde{V}E=V$ and $\tilde{W}F=W$.  Together with their conjugates, these equations imply $[\tilde{V},E]=0=[\tilde{W},F]$, and hence that 
\[ \tilde{V} = V + (1-E)\tilde{V}(1-E),~~~~~\tilde{W}=W + (1-F)\tilde{W}(1-F) . \]
Substitution into  
\[
(\tilde{V}-\langle y_j|\rho|y_j\rangle)\,|x_j\rangle = 0,~~~~~(\tilde{W}-\langle x_j|\rho|x_j\rangle )|y_j\rangle =0 ,
\]
(which is equivalent to Eq.~(\ref{prop2a}) precisely as per the equivalence of Eqs.~(\ref{prop1a}) and (\ref{prop1b}) in Proposition 1), and using Eqs.~(\ref{prop1a}) and (\ref{aex}), then gives
\[ (1-E)\tilde{V}(1-E)|x_j\rangle = 0 = (1-F)\tilde{W}(1-F)|y_j\rangle \]
for all $j$.  But $\{|x_j\rangle\}$ and $\{|y_j\rangle\}$ are basis sets for $H_\infty$, implying the operators must vanish identically, and the desired result immediately follows.


\begin{thebibliography}{99}
\bibitem{footnote} Note that the coincidence rate reaches the maximum possible value of unity for $\alpha=1$, with $A^{(1)}$ and $B^{(1)}$ corresponding to orthogonal POMs generated by the Schmidt basis $\{ |\phi^{(1)}_2\rangle\otimes|\phi^{(1)}_2\rangle, |\phi^{(1)}_3\rangle\otimes|\phi^{(1)}_3\rangle\}$ for $|\psi\rangle$.
\end{thebibliography}
\end{document}